\documentclass[aps,prx,twocolumn,nofootinbib,superscriptaddress,10pt]{revtex4-2}

\usepackage{graphicx}
\usepackage{amsmath,amssymb,amsfonts, mathtools}
\usepackage{mathrsfs}
\usepackage{bm}
\usepackage{enumitem}
\usepackage{hyperref}
\usepackage{xcolor}
\usepackage{microtype}

\DeclareMathOperator{\diag}{diag}

\hypersetup{
    colorlinks=true,
    linkcolor=blue,
    filecolor=magenta,      
    urlcolor=blue,
    citecolor=blue,
}

\begin{document}

\title{Learning as a Geometric Phase Transition: \texorpdfstring{\\}{ } Renormalization Group Flow and Anisotropic Symmetry Breaking in Deep Networks}

\author{Giacomo Le Pera}
\affiliation{Independent Researcher}
\author{Claudio Nordio}
\affiliation{Independent Researcher}

\date{\today}

\begin{abstract}
We formulate feature learning as a geometric critical phenomenon of the lifted tensor-product learning metric. The central object is not a scalar overlap, but the target-active geometry of
\[
\mathcal N_{0,L}=\frac1N\sum_{r=1}^{L}\Sigma_{r\to L}\otimes T_{0\to r-1},
\]
which entangles forward pullback survival with backward push-forward visibility. The neutral phase is target-isotropic: after restriction to endpoint target-active states and trace normalization, the lifted metric is proportional to the identity. Learning corresponds to an instability of this target-isotropic fixed point and to the emergence of traceless target-aligned eigentensors. We derive discrete Dyson expansions for local anisotropic insertions and their continuous Callan--Symanzik flow. Crucially, before constructing the full temporal mean-field theory, we identify the local spatial source of the $\beta$-functions directly from microscopic kinematics: asynchronous gradient updates generate synchronous metric strains, whose target-active symmetric traceless components act as curvature-like defects. The Wilsonian depth RG flow is then governed by the transport, balance, and coarse-grained irrelevance of these defects. Heavy-tailed spectra arise, under a scale-free counting hypothesis, as the spectrum of the target-active lifted geometry, with exponent addition in the matched pullback--push-forward sector. Finally, we relate this depth RG picture to temporal stochastic training dynamics and to the kinematic imprint of the learned channel on empirical weight Gram matrices.
\end{abstract}

\maketitle

\section{Collective closure and induced metric transport}

Following \cite{Nordio}, consider a feed-forward ReLU network trained by gradient descent. At layer $\ell$ let
\[
\nu_\ell^\alpha=W^{(\ell)}u_{\ell-1}^\alpha,
\qquad
\nu_\ell^\alpha\mapsto u_\ell^\alpha=\phi(\nu_\ell^\alpha),
\qquad
A_\ell^\alpha=\diag(\phi'(\nu_\ell^\alpha)),
\]
and let $b_\ell^\alpha$ denote the conjugate backward field associated with sample $\alpha$. The standard residual update has the schematic form
\[
R(t+1)=R(t)-\frac{2\eta}{M}K(t)R(t),
\]
where the learning kernel $K$ is obtained by contracting endpoint forward and backward tangent states. For shallow networks, the collective dynamics can often be expressed using only forward overlaps $Q$ and backward overlaps $S$. For depth $L\ge3$, however, the update of $b_\ell$ contains intermediate weight-induced transports. Markovian closure in collective variables therefore forces the introduction of induced Gram metrics. A typical forward pullback metric is
\[
G_\ell^{\alpha}=\big(W^{(\ell+1)}\big)^T A_{\ell+1}^{\alpha}W^{(\ell+1)},
\]
and the corresponding backward visibility metric is
\[
\widetilde G_\ell^{\alpha}=A_\ell^\alpha W^{(\ell)}\big(W^{(\ell)}\big)^T A_\ell^\alpha .
\]
These objects are not auxiliary decoration: they are the curvilinear transports that make the backward--forward dynamics Markovian at the collective level. Thus, from depth three onward, learning is naturally described as a dynamics of induced metrics rather than only as a dynamics of weights.

\section{The lifted metric and the depth RG channel}

In the following, we adopt approach and definitions given in \cite{Nordio}, introducing a further level of abstraction. For each layer $r$, define a bidirectional learning channel
\[
\boxed{
\mathcal C_r^{\alpha\beta}
=
\Sigma_{r\to L}^{\alpha\beta}\otimes T_{0\to r-1}^{\alpha\beta} .
}
\]
Here $T_{0\to r-1}^{\alpha\beta}$ is the forward pullback metric measuring which input-side directions survive up to layer $r-1$, while $\Sigma_{r\to L}^{\alpha\beta}$ is the backward push-forward metric measuring which directions at layer $r$ are visible from the readout and residual flow. The lifted tensor-product learning metric is
\[
\boxed{
\mathcal N_{0,L}^{\alpha\beta}
=
\frac1N\sum_{r=1}^{L}
\mathcal C_r^{\alpha\beta}
=
\frac1N\sum_{r=1}^{L}
\Sigma_{r\to L}^{\alpha\beta}\otimes T_{0\to r-1}^{\alpha\beta} .
}
\]
The ordinary scalar learning kernel is obtained by evaluating this lifted operator on endpoint tangent states. For a scalar readout $f^\alpha=b_L^T u_L^\alpha$, the endpoint covector $b_L$ is sample-independent; the sample dependence of the local backward fields is carried by the transported visibility operators $\Sigma_{r\to L}^{\alpha\beta}$. Thus
\[
\zeta^\alpha=b_L\otimes u_0^\alpha,
\]
and schematically
\[
K_{\alpha\beta}^{(L)}
=
(\zeta^\alpha)^T\mathcal N_{0,L}^{\alpha\beta}\zeta^\beta .
\]
The lifted metric contains strictly more information than $K$: it remembers which learnability modes arise from forward survival, which arise from backward visibility, and which arise from genuine pullback--push-forward entanglement. We introduce $\mathcal N$ to locate the scalar kernel inside the larger bidirectional geometry. In the remainder of this paper we focus on the layerwise channel $\mathcal C_r$, which is the local object whose deformation directly drives the depth RG flow. A more general treatment of the normalized target-active geometry of $\mathcal N$ as an order parameter, together with the associated Wilsonian recursion for relevant lifted eigentensors, is deferred to future work.

We treat the depth coordinate as an RG coordinate.  Adding a layer or coarse-graining a block of layers maps the normalized target-active geometry of the channel into a new one. The core question is whether this map only rescales an already existing channel, or whether it produces new target-aligned anisotropic structure. The former is pleonastic depth; the latter is feature learning.

\paragraph*{Scope and assumptions.}
The derivations below are intended as a controlled scaling theory. We work in a coarse-grained regime where ReLU chambers are locally stable or averaged over, isotropic backgrounds can be separated from anisotropic insertions, target-active spectral modes are isolated from the dense bulk, and accumulated gradient updates are coherent along task-aligned directions. These assumptions are stated explicitly whenever they are used. We also work in the ReLU/fixed-chamber transport class, where state and tangent transports coincide. For smooth non-homogeneous activations this degeneracy is broken, we believe in a tractable way \cite{Nordio&LePera} as a perturbation of the present lifted-channel fixed point.

\section{Anisotropic Symmetry Breaking and Discrete Dyson Equations}
At the neutral fixed point, the channel exhibits an effective target-isotropy: each layer acts, to leading order, as a partial isometry on the target-active subspace. For a layer with Jacobian $J_r = A_r W^{(r)}$, the forward metric and the backward visibility metric satisfy:
\begin{align}
    (W^{(r)})^T A_r W^{(r)} &\approx \tau_r I \\
    A_r W^{(r)} (W^{(r)})^T A_r &\approx \tau_r A_r
\end{align}
where $\tau_r$ is a scalar scale factor. In this state, the layer is metric-neutral.

Feature learning corresponds to a target-induced anisotropic instability of this neutral geometry. The gradient aligns the weights to the target, creating an anisotropic perturbation. We formalize this locally by writing the metrics as a background isotropic scalar plus a target-aligned operator (denoted perturbatively as $\epsilon \Omega$):
\begin{align}
    (J_r)^T J_r &= \tau_r I + \epsilon \Omega_r^F \\
    J_r (J_r)^T &= \tau_r A_r + \epsilon \Omega_r^B
\end{align}
where $\Omega_r^F$ and $\Omega_r^B$ contain the target-aligned anisotropies. Using this ansatz, a forward metric recursion can be expanded perturbatively. In the scalar-background approximation, or after transporting local insertions to a common coarse-grained frame, such recursion simplifies in $T_{0 \to r} = (J_r)^T J_r T_{0 \to r-1}$ and the first-order target-aligned correction is:
\begin{equation}
    T^{(1)}_{0 \to r} = \left( \prod_{m=1}^r \tau_m \right) \sum_{j=1}^r \frac{1}{\tau_j} \Omega_j^F
\end{equation}
This is the discrete Dyson form used throughout the paper. It reveals that the geometric deformation of the learning channel is a weighted sum of symmetry-breaking insertions $\Omega_j^F$. Crucially, the factor $1/\tau_j$ shows that bottlenecks (layers with small isotropic scales) exert disproportionate influence over the global geometry, because their anisotropic insertion is not suppressed by their own scalar transmission.

\subsection{The Jacobian as a Geometric Propagator}
To fully grasp the physical meaning of the Renormalization Group flow, it is instructive to map our geometric operators to the standard formalism of statistical mechanics and Quantum Field Theory. In this analogy, the masked Jacobian $J_r = A_r W^{(r)}$ acts exactly as a local discrete-time propagator, or transfer matrix, along the depth coordinate of the network.

Forward activations evolve via $u_r = J_r u_{r-1}$, while the conjugate error fields evolve backward via the adjoint operator $b_{r-1} = (J_r)^T b_r$. The macroscopic transport operator $F_{0 \to r} = J_r J_{r-1} \dots J_1$ thus serves as the discrete Green's function of the network, dictating how a localized input perturbation propagates to layer $r$.

In quantum mechanics, a probability amplitude $\psi$ is not directly observable; the physically measurable quantity is its squared modulus $\psi^\dagger \psi$. Our geometric framework perfectly mirrors this structure. The Jacobians $J_r$ and the propagator $F$ are linear maps between distinct vector spaces (from layer $r-1$ to layer $r$), and thus cannot act as distance-measuring metrics themselves. To construct a physically measurable geometric invariant, the propagator must be "squared" against its adjoint. This operation recovers precisely the forward pullback metric $T = F^T F$ and the backward push-forward metric $\Sigma = F D F^T$.

Consequently, applying the continuous depth limit to these operators is mathematically equivalent to performing a Renormalization Group flow directly on the network's macroscopic propagator. The symmetry-breaking anisotropies $\omega_F$ and $\omega_B$ act as dynamic mass or coupling terms that accumulate on the propagator space. As training progresses, this accumulation can drive the system toward a critical regime where the geometric observables obey Callan--Symanzik-type equations and condense into scale-invariant power laws.

\section{Continuous Limit and Callan-Symanzik Equation}
We promote the depth index to a continuous variable $r \in [0,L]$. This should be understood as a thermodynamic mean-field limit in depth: the continuous coarse-graining effectively smooths over the discrete, non-differentiable mask-switching events of individual ReLU gates, capturing the macroscopic geometric flow of the network.

We define the local densities $\tau_r \approx 1 + \gamma(r)dr$ and $\epsilon \Omega_r^F \approx \omega_F(r)dr$. The ordered product of layer transports converges to a path-ordered exponential.  We will not need its full non-commutative form here; at the level of the coarse-grained generator the continuous flow of the forward metric obeys:
\begin{equation}
    \frac{\partial T(r)}{\partial r} = \left[ \gamma(r)I + \omega_F(r) \right] T(r)
\end{equation}
Applying the same procedure to the backward visibility $\Sigma(r)$, we derive the flow for the total learning channel $\mathcal{C}(r) = \Sigma(r) \otimes T(r)$. By Leibniz's rule, this yields a continuous RG equation analogous to the Callan-Symanzik equation:
\begin{align}
    \frac{\partial \mathcal{C}(r)}{\partial r} &= \Sigma(r) \otimes \left( [\gamma(r)I + \omega_F(r)] T(r) \right) \nonumber \\
    &- \left( [\tilde{\gamma}(r)I + \omega_B(r)] \Sigma(r) \right) \otimes T(r)
\end{align}

At this stage $\omega_F$ and $\omega_B$ should be read as effective anisotropic generators of the forward and backward metric transports. We first analyze the RG consequences of such generators for the normalized lifted channel. Only later, in the microscopic kinematic section, do we identify their local origin in the synchronous metric strains produced by gradient descent. Thus the paper proceeds from an effective RG description to a local pre-DMFT closure of its anisotropic source.

\section{Dynamic Feedback and Emergence of Power Laws}
To understand how the system reaches a non-trivial fixed point, we must isolate the geometric shape of the learning channel from its overall magnitude. Let $P_{\mathcal{T}}$ denote the projection onto the target-relevant endpoint states. We define the partition trace $Z(r) = \operatorname{Tr}_{\mathcal T}(P_{\mathcal{T}} \mathcal{C}(r) P_{\mathcal{T}})$ and the normalized tensor $\hat{\mathcal{C}}(r) = \mathcal{C}(r)/Z(r)$.

The scale-free geometry evolves according to the tensorial $\beta$-function:
\begin{equation}
    \beta(\hat{\mathcal{C}}) = r \frac{\partial \hat{\mathcal{C}}(r)}{\partial r}
\end{equation}
By substituting the Callan--Symanzik equation and expanding the derivative of the quotient, the purely isotropic terms ($\gamma, \tilde{\gamma}$) factor out and cancel with the trace (as shown in Appendix A). The $\beta$-function thus depends on the anisotropic insertions $\omega_F$ and $\omega_B$. If $\omega_F = \omega_B = 0$, $\beta(\hat{\mathcal{C}}) = 0$, trapping the network in the trivial pleonastic fixed point.

\subsection{The Spectral Beta Function and Dual Stationarity}
To solve $\beta(\hat{\mathcal{C}}) = 0$ outside the trivial regime, we project the operator onto its target-active eigenvectors. We specifically restrict our analysis to the isolated, task-relevant leading modes (spectral outliers). For these modes, the eigengaps separating them from the dense bulk are assumed sufficiently large to neglect second-order eigenvector rotations. This makes the scalar spectral evolution a controlled leading-order approximation.

Let $\hat{\nu}_k(r)$ be the eigenvalues of the normalized channel along these isolated modes. The condition for a fixed point translates to the \textit{spectral Beta function} vanishing for all relevant modes $k$:
\begin{equation}
    \beta_k = r \frac{\partial \log \hat{\nu}_k(r)}{\partial r} \approx 0
\end{equation}
Here the effective anisotropic generators $\omega_F$ and $\omega_B$ are dynamic: they are generated by accumulated gradient updates, which are proportional to the current error residual $r(t)$. Therefore, the emergence of the non-trivial fixed point is described by the compatibility of two stationarity notions:
\begin{itemize}
    \item \textbf{Temporal Stationarity:} Gradient descent minimizes the loss until the residual vanishes along the active directions of the kernel, $K_* r_* \simeq 0$.
    \item \textbf{Depth Stationarity:} As the error $r(t)$ decays, the forcing terms $\omega_F, \omega_B$ cease to grow and settle on asymptotic values $\omega_*, \tilde{\omega}_*$. At this equilibrium, the normalized channel satisfies $\hat{\mathcal{C}}_{r+1,*} \approx \hat{\mathcal{C}}_{r,*}$.
\end{itemize}
Temporal stationarity alone does not imply a spatial RG fixed point. It freezes the learned depth profile. The geometric phase transition corresponds to the additional condition that this frozen profile is stable under depth coarse-graining. The flow reaches a non-trivial attractive fixed point when the network deforms its channel geometry enough to align the relevant eigenvectors with the target; once this alignment is achieved, the residual drops, the driving terms cease to grow, and the normalized tensorial $\beta$-function can approach zero.

\subsection{Critical Power Laws and Rank-Index Definition}
At a scale-invariant critical fixed point ($\beta_k = 0$), all non-normalized target-active eigenvalues $\nu_k(r)$ change by the same scale factor across depth. We write this scaling ansatz as:
\begin{equation}
    \frac{\partial \nu_k(r)}{\partial r} = - \frac{z_N}{r} \nu_k(r)
\end{equation}
where $z_N\equiv z_{\mathcal N}$ is the scaling dimension of the lifted learning channel in depth. Integrating this equation yields a pure scaling law under depth reparameterization $r \mapsto br$:
\begin{equation}
    \nu_k(b r) \simeq b^{-z_N} \nu_k(r)
\end{equation}

The vanishing of the normalized beta function gives stationarity of the target-active shape. A power-law spectrum follows once this stationary shape is assumed to be scale invariant in the sense of an effective counting dimension. At such a scale-invariant fixed point, the natural spectral scaling hypothesis gives a critical power law for the target-active learning spectrum. If the effective number of active modes grows as $n \mapsto b^{d_{\text{eff}}} n$, the integrated spectral counting function---which counts the number of modes whose eigenvalue is strictly greater than a threshold $\nu$---satisfies:
\begin{equation}
    N_{>}(\nu) = \#\{k : \nu_k > \nu\} \sim \nu^{-d_{\text{eff}}/z_N}
\end{equation}
We can invert this relation to describe the ordered spectrum. Let $a \in \{1, 2, \dots, M\}$ be the \textbf{rank index} of the eigenvalues sorted in descending order (where $a=1$ identifies the largest eigenvalue, representing the most open and receptive learning direction). The relation yields the rank-ordered geometric power law:
\begin{equation} \label{eq:spatial_law}
    \nu_a \sim a^{-p_N}, \quad \text{with} \quad p_N = \frac{z_N}{d_{\text{eff}}}
\end{equation}
Consequently, the macroscopic condensation of the learning spectra and their heavy-tailed distributions are recovered as a scaling signature of a backward-forward RG fixed point, rather than as a purely statistical artifact.

\subsection{Matched Pullback--Push-Forward Exponent Addition}
The previous scaling law concerns the target-active spectrum of the learning channel. It is important to distinguish this spectrum from the full bulk spectrum of the tensor product. Suppose, in a common-mode approximation, that the forward and backward metrics have matched modes
\[
T f_k \simeq \tau_k f_k,
\qquad
\Sigma e_k \simeq \sigma_k e_k,
\]
with power laws
\[
\tau_k\sim k^{-p_F},
\qquad
\sigma_k\sim k^{-p_B}.
\]
On the matched lifted mode $e_k\otimes f_k$ one has
\[
(\Sigma\otimes T)(e_k\otimes f_k)
\simeq
(\sigma_k\tau_k)(e_k\otimes f_k),
\]
and therefore
\begin{equation}
    \nu_k\sim \sigma_k\tau_k\sim k^{-(p_F+p_B)}.
\end{equation}
Thus the matched learnability exponent is
\begin{equation}
    p_{\rm learn}=p_F+p_B.
\end{equation}
This addition law should not be confused with the exponent of the full bulk tensor spectrum, which contains all products $\sigma_i\tau_j$. The exponent addition is a signature of target-active matching: the target correlates, or entangles in the lifted channel, a forward-surviving direction with a backward-visible direction.

\subsection{Kinematic Imprint: Why the Weight Gram Matrix Inherits the Power Law}
While the RG critical fixed point predicts a power-law spectrum $\nu_a \sim a^{-p_N}$ for the geometric learning channel $\mathcal{C}_r$, empirical observations report heavy-tailed spectral condensation on the ``naked'' weight Gram matrices $(W^{(r)})^T W^{(r)}$ \cite{Martin2021}. We now give the kinematic mechanism by which the channel spectrum can be imprinted on the empirical weight Gram spectrum.

The gradient update for the weight matrix at layer $r$ is driven by the outer product of the forward activations $u_{r-1}$ and the masked backward conjugate fields $c_r = A_r b_r$:
\begin{equation}
    \Delta W^{(r)} = - \frac{2\eta}{MN} \sum_{\beta=1}^M r^\beta c_r^\beta (u_{r-1}^\beta)^T
\end{equation}
To determine how this update shapes the geometric space, we can test it against a rank-one perturbation $pq^T$ in the local parameter space of $W^{(r)}$ using the Frobenius inner product:
\begin{equation}
    \langle \Delta W^{(r)}, pq^T \rangle_F = - \frac{2\eta}{MN} \sum_{\beta=1}^M r^\beta (p^T c_r^\beta) (q^T u_{r-1}^\beta)
\end{equation}
This shows that layer-wise learnability requires the simultaneous alignment of a forward component $(q^T u_{r-1}^\beta)$ and a backward component $(p^T c_r^\beta)$, bridged by the target residual $r^\beta$. The spatial covariance of the update is therefore controlled by the same forward pullback and backward push-forward geometries that define the bidirectional channel $\mathcal{C}_r$.

As the network flows toward the critical RG fixed point, the target-active spectrum of $\mathcal{C}_r$ condenses into a scale-invariant power law. Since the final weight matrix $W^{(r)}(t)$ is the temporal integral of these updates, it can coherently accumulate mass along the heavy-tailed eigenmodes selected by $\mathcal{C}_r$, provided the target-aligned updates dominate the isotropic initialization and temporal cross-terms do not wash out the signal. Under this coherence assumption, the heavy-tailed spectrum observed in $(W^{(r)})^T W^{(r)}$ is predicted to appear as the kinematic imprint of the geometric phase transition.

\subsection{Unifying Time and Space: Kesten Dynamics and the Optimal Exponent \texorpdfstring{$\alpha=2$}{alpha=2}}
The kinematic link between the abstract channel $\mathcal{C}_r$ and the weight Gram matrix provides a dynamical interpretation of empirical heavy-tailed spectra. Empirical spectral analyses report that deep networks often generalize best when the spectral density of weight matrices follows a power law $\rho(\nu) \sim \nu^{-\alpha}$ with $\alpha$ near $2$ \cite{Martin2021}. Our geometric theory interprets this observation as a compatibility condition between the temporal SGD dynamics and the spatial depth-wise RG flow.

This subsection is not a derivation of a universal exponent. It is a minimal phenomenological compatibility argument between temporal multiplicative dynamics and the spatial RG spectrum.

How do the temporal and spatial power laws talk to each other? The temporal training dynamics act as the microscopic driver pushing the network toward its macroscopic spatial (RG) critical point. A minimal phenomenological model for the active eigenvalues is a Kesten-type multiplicative recursion in time,
\[
\lambda_{t+1}=A_t\lambda_t+B_t .
\]

We parameterize the multiplier as $A_t = 1 - \kappa_t + 2\chi_t$. Here, $-\kappa_t$ represents the deterministic drift contracting the eigenmode as the residual is dissipated. The term $\chi_t$ is the normalized random increment induced by state-dependent mini-batch sampling, making the noise multiplicative. $B_t$ acts as a positive reinjection term. 

Under these conditions, if the log-multiplier $\log A_t$ is approximately Gaussian with mean $m < 0$ and variance $v > 0$---a mean-field assumption appropriate to the asymptotic late-training basin, where averaged mini-batch fluctuations are effectively coarse-grained---the stationary spectral density develops a temporal Kesten power-law tail:
\begin{equation} \label{eq:temporal_density}
    \rho(\nu) \sim \nu^{-(1+p)}, \quad \text{where } p = -\frac{2m}{v}
\end{equation}
This temporal density should be compatible with the spatial RG geometry. If the probability density scales as $\nu^{-(1+p)}$, the cumulative rank index scales as $a \sim \nu^{-p}$. Inverting this relationship, the rank-ordered eigenvalues produced by the temporal Kesten process scale as:
\begin{equation}
    \nu_a \sim a^{-1/p}
\end{equation}
By equating this temporal exponent with the spatial RG exponent derived in Eq. \eqref{eq:spatial_law}, we find the unification condition:
\begin{equation}
    p_N = \frac{1}{p}
\end{equation}

Equating the phenomenological exponent $\alpha = 2$ to our theoretical density $(1+p) = 2$ yields $p = 1$, and therefore a geometric RG exponent $p_N = 1$. The physical significance of this value is simple. The condition for the Kesten power law is $\mathbb{E}[A_t^p] = 1$ \cite{Kesten1973}. At $p=1$, this imposes marginal balance: $\mathbb{E}[A_t] = 1$. In terms of logarithmic increments, substituting $p=1$ yields:
\begin{equation}
    m = - \frac{v}{2}
\end{equation}
This is the signature of a geometric Brownian motion at marginal criticality. The deterministic contraction exerted by the gradient descent drift ($m$) balances the expansive stochastic fluctuations introduced by the mini-batch sampling ($v/2$). This gives a dynamical reading of the generalization boundaries observed in empirical spectral analyses \cite{Martin2021}:
\begin{itemize}
    \item \textbf{Overfitting ($\alpha < 2 \implies p < 1$):} We find $m > -v/2$. The stochastic variance dominates, injecting excessive mass into isolated eigenmodes and causing the network to memorize noise.
    \item \textbf{Underfitting ($\alpha > 2 \implies p > 1$):} We find $m < -v/2$. The deterministic decay dominates, suppressing the formation of target-aligned heavy tails and trapping the network in a nearly isotropic state.
\end{itemize}
Thus, within the lognormal Kesten approximation, the empirically important exponent $\alpha = 2$ corresponds to stationary marginal dynamics in time, compatible with a non-trivial spatial RG fixed point in depth.

\section{Microscopic Kinematics and Wilsonian Depth RG}

Before carrying out a full open-tensor DMFT program, it is highly instructive to isolate the local mechanism by which gradient descent breaks target-isotropy. By focusing on the geometric transport of the local metric strain, we can identify the local spatial source of the Callan--Symanzik flow directly from kinematics, before closing the full path-integral loop.

\subsection{From Asynchronous Weight Updates to Curvature-Like Metric Insertions}
The elementary gradient step at a layer is inherently asynchronous: it maps the incoming feature space to the outgoing feature space. In a fixed ReLU chamber, the layer Jacobian is $J_r=A_r W^{(r)}$, and the first-order weight update has the schematic rank-one form
\[
\Delta W_r \sim -\eta\sum_\beta \Delta^\beta \,(A_r^\beta b_r^\beta)(u_{r-1}^\beta)^T ,
\]
where we now denote the target residual as $\Delta^\beta$ to avoid confusion with the layer coordinate $r$. Thus
\begin{equation}
\boxed{ \Delta W_r \sim |b_r\rangle\langle u_{r-1}| }
\end{equation}
up to masks, residual weights, and sample summation. This object is asynchronous because it lives between two different layer spaces, $\Delta W_r: V_{r-1} \longrightarrow V_r$. It is therefore best viewed as a local variation of transport, or connection, rather than as a metric deformation by itself.

The anisotropic geometry appears only after one leg is closed with the current transport. For the input-side local Gram, $G_r^- := J_r^T J_r$, its variation is
\[
\Delta G_r^- = \Delta J_r^T J_r + J_r^T \Delta J_r + O(\eta^2).
\]
Using the rank-one structure of $\Delta J_r$, this produces synchronous endomorphisms on $V_{r-1}$:
\[
J_r^T\Delta J_r \sim |b_{r-1}\rangle\langle u_{r-1}|, \qquad \Delta J_r^T J_r \sim |u_{r-1}\rangle\langle b_{r-1}|.
\]
Hence
\begin{equation}
\boxed{ \Delta(J_r^T J_r) \sim |u_{r-1}\rangle\langle b_{r-1}| + |b_{r-1}\rangle\langle u_{r-1}| }
\end{equation}
Dually, the output-side local Gram $G_r^+ := J_r J_r^T$ receives the synchronous contribution on $V_r$:
\begin{equation}
\boxed{ \Delta(J_r J_r^T) \sim |u_r\rangle\langle b_r| + |b_r\rangle\langle u_r| }
\end{equation}
Thus the gradient step is asynchronous at the level of weights, but becomes synchronous once one leg is closed by the current layer transport. These synchronous rank-one endomorphisms are the true local sources of metric anisotropy.

This suggests a connection-like interpretation. The update $\Delta J_r$ is a variation of the local transport. Closing one leg gives $\Omega_r^- := J_r^T \Delta J_r$, such that $\Delta G_r^- = \Omega_r^- + (\Omega_r^-)^T + O(\eta^2)$. 

The local synchronous endomorphism $E_r$ naturally decomposes into three sectors:
\[
E_r = \text{trace} + \text{antisymmetric} + \text{symmetric traceless}.
\]
Under depth coarse-graining, these sectors possess distinct physical meanings:
\begin{align*}
\text{trace} &\longrightarrow \text{isotropic scale} \\
\text{antisymmetric} &\longrightarrow \text{frame (gauge) rotation} \\
\text{symmetric traceless} &\longrightarrow \text{candidate relevant operator}
\end{align*}
The antisymmetric part of $\Omega_r^-$ rotates the frame and is gauge-like; the trace rescales the local metric isotropically; the symmetric traceless part changes the geometric \textit{shape}. Therefore, the local anisotropic source is naturally:
\[
\operatorname{Sym}_0(\Omega_r^-) = \operatorname{Sym}(\Omega_r^-) - \frac{\operatorname{Tr} \operatorname{Sym}(\Omega_r^-)}{d_r} I_{r-1}.
\]
After projection onto a target- or residual-active probe subspace $P_{\mathcal T}$, the natural local curvature-like insertions are:
\begin{align}
\boxed{ \mathfrak F_r^- = P_{\mathcal T} \operatorname{Sym}_0 \left[ \sum_\beta \Delta^\beta |u_{r-1}^\beta\rangle\langle b_{r-1}^\beta| \right] P_{\mathcal T} } \\
\boxed{ \mathfrak F_r^+ = P_{\mathcal T} \operatorname{Sym}_0 \left[ \sum_\beta \Delta^\beta |u_r^\beta\rangle\langle b_r^\beta| \right] P_{\mathcal T} }
\end{align}
These are curvature-like because they measure the failure of the training-time deformation and depth transport to commute: an infinitesimal move in training time changes the metric seen by an infinitesimal move in depth. 

\subsection{Applying Flatness to the Local Anisotropic Sources}
The lifted Callan--Symanzik source can then be read locally by identifying $\omega_F \sim \mathfrak F_r^-$ and $\omega_B \sim \mathfrak F_r^+$. The corresponding layerwise shape source is therefore:
\begin{equation}
\boxed{ \mathscr S_r \simeq \Sigma_{r\to L}\otimes (\mathfrak F_r^- T_{0\to r-1}) - (\mathfrak F_r^+\Sigma_{r\to L}) \otimes T_{0\to r-1} }
\end{equation}
At the lifted fixed point, one should not require these curvature-like insertions to vanish absolutely. The learned phase may have a nonzero anisotropic geometry. What vanishes is the shape flow of the \textit{normalized} target-active channel. The flatness condition is not $\mathscr S_r = 0$ as a raw operator. Rather, after target-active projection and trace removal, it requires:
\begin{equation}
\boxed{ \Pi_{\mathcal T,0} [\mathscr S_r] = 0 }
\end{equation}
where $\Pi_{\mathcal T,0}$ denotes target-active projection followed by the removal of the isotropic trace component. Thus, the fixed point is flat only in the normalized anisotropic sector.

There is also an equivalent local transport form. Since $u_r = J_r u_{r-1}$ and $b_{r-1} = J_r^T b_r$, the two endomorphisms $E_r^-$ and $E_r^+$ are the \textit{same} rank-one strain viewed on the two sides of the layer. If the layer is locally metric-neutral up to scale ($J_r^T J_r \simeq \tau_r I$), then the incoming and outgoing anisotropic sources should agree after parallel transport, modulo trace and gauge rotations. This defines the local curvature defect on the outgoing side:
\begin{equation}
\boxed{ \mathcal K_r^+ = \operatorname{Sym}_0\left[ \mathfrak F_r^+ - \tau_r^{-1} J_r \mathfrak F_r^- J_r^T \right] }
\end{equation}
The local flatness condition is then $P_{\mathcal T}\mathcal K_r P_{\mathcal T} = 0$. In words: flatness means that the anisotropic strain generated by the gradient step is covariantly constant across the layer.

\subsection{Iterating the Flatness Condition (Wilsonian RG)}
The local flatness condition can be iterated to provide a direct route to a depth coarse-graining map. The key physical observation is the continuity of the strain: the outgoing synchronous endomorphism of layer $r$ is exactly the incoming synchronous endomorphism of layer $r+1$:
\begin{equation}
E_r^+ = \sum_\beta \Delta^\beta |u_r^\beta\rangle\langle b_r^\beta| = E_{r+1}^-
\end{equation}
Thus, up to the variation of the target-active projectors, $\mathfrak F_r^+ \simeq \mathfrak F_{r+1}^-$.

Combining this continuity with the local flatness relation $\mathfrak F_r^+ \simeq \Pi_{r,\mathcal T,0} [\tau_r^{-1}J_r\mathfrak F_r^- J_r^T]$, we obtain the fundamental layer-to-layer recursion:
\begin{equation}
\boxed{ \mathfrak F_{r+1}^- \simeq \mathcal R_r[\mathfrak F_r^-], \qquad \mathcal R_r[X] = \Pi_{r,\mathcal T,0} \left[\tau_r^{-1} J_r X J_r^T\right] }
\end{equation}
Iterating over $k$ layers in an autosimilar region ($\mathcal R_r \simeq \mathcal R_*$) yields $\mathfrak F_{r+k}^- \simeq \mathcal R_*^k[\mathfrak F_r^-]$.

The relevant geometric operators are then simply the eigentensors of this coarse-grained transport: $\mathcal R_*[\mathcal O_a] = \lambda_a \mathcal O_a$. If we decompose the incoming strain as $\mathfrak F_r^- = \sum_a g_a(r)\mathcal O_a$, we find:
\begin{equation}
\boxed{ g_a(r+k) \simeq \lambda_a^k g_a(r) }
\end{equation}
Thus, $\lambda_a > 1$ identifies a relevant anisotropic operator, $\lambda_a < 1$ an irrelevant one, and $\lambda_a = 1$ a marginal one. In this formulation, the symmetry-breaking direction is an eigenoperator of the depth-transported strain.

In a continuous-depth approximation, writing $J_r \simeq I + ds\,\Gamma(s)$ and $\tau_r = 1 + ds\,\gamma(s)$, the transport recursion $\tau_r^{-1} J_r X J_r^T \simeq X + ds\,[\Gamma X + X\Gamma^T - \gamma X]$ yields a local Callan--Symanzik-type equation for the anisotropic source:
\begin{equation}
\boxed{ \partial_s\mathfrak F = \Pi_{\mathcal T,0} \left[ \Gamma\mathfrak F + \mathfrak F\Gamma^T - \gamma\mathfrak F \right] + \mathcal K(s) }
\end{equation}
At a flat fixed point, the curvature defect $\mathcal K$ is irrelevant and $\partial_s\mathfrak F = 0$. This confirms our geometric intuition: the fixed point is not the absence of anisotropy, but rather \textit{stable anisotropy}, where the target-selected symmetric traceless strain is transported coherently from layer to layer along the depth manifold.

\subsection{Deriving the Coarse-Grained Beta Functions from Strain Mismatch}
We can now express the local source of the macroscopic tensorial and spectral $\beta$-functions directly in terms of the microscopic synchronous strains. This is a local pre-DMFT kinematic closure: it identifies the source that enters the depth RG flow, while the full temporal mean-field theory remains needed to determine the statistics, correlations, and self-consistency of these strains across training time and depth.

Substituting the kinematically derived local strains into the lifted source gives
\begin{equation}
    \mathscr{S}_r = \Sigma_{r \to L} \otimes (\mathfrak{F}_r^- T_{0 \to r-1}) - (\mathfrak{F}_r^+ \Sigma_{r \to L}) \otimes T_{0 \to r-1} .
\end{equation}
The target-active tensorial $\beta$-function is therefore determined, at leading order in the local strain insertions, by the centered anisotropic source
\begin{equation}
    \boxed{
    \beta(\hat{\mathcal{C}}_{\mathcal T})
    = r\frac{\partial \hat{\mathcal{C}}_{\mathcal T}}{\partial r}
    = r\left[
    \frac{P_{\mathcal T}\mathscr{S}_rP_{\mathcal T}}{Z}
    -\hat{\mathcal{C}}_{\mathcal T}
    \operatorname{Tr}_{\mathcal T}\left(\frac{P_{\mathcal T}\mathscr{S}_rP_{\mathcal T}}{Z}\right)
    \right] .
    }
\end{equation}
Thus the raw source does not need to vanish at a geometric RG fixed point. What must vanish is its shape-changing part after trace normalization. Equivalently, within the relevant target-active subspace, the source may survive only as a global dilation of the channel, which is removed by the normalized flow.

To see the physical constraint imposed on individual modes, project this tensorial beta function onto an isolated matched target-active eigenmode
\[
    v_k = |\psi_B\rangle\otimes |\psi_F\rangle,
\]
with unnormalized eigenvalue $\nu_k=\sigma_k\tau_k$, where
\[
    \Sigma |\psi_B\rangle = \sigma_k |\psi_B\rangle,
    \qquad
    T |\psi_F\rangle = \tau_k |\psi_F\rangle .
\]
The isolated-mode assumption allows a Hellmann--Feynman projection of the leading eigenvalue flow. Evaluating the raw source gives
\begin{align}
    \langle v_k | \mathscr{S}_r | v_k \rangle
    &= \langle \psi_B | \Sigma | \psi_B \rangle
       \langle \psi_F | \mathfrak{F}_r^- T | \psi_F \rangle \nonumber \\
    &\quad - \langle \psi_B | \mathfrak{F}_r^+ \Sigma | \psi_B \rangle
       \langle \psi_F | T | \psi_F \rangle .
\end{align}
Define the scalar strains seen by the matched forward and backward legs as
\begin{equation}
    f_k^- \equiv \langle \psi_F | \mathfrak{F}_r^- | \psi_F \rangle,
    \qquad
    f_k^+ \equiv \langle \psi_B | \mathfrak{F}_r^+ | \psi_B \rangle .
\end{equation}
Then
\begin{align}
    \langle v_k | \mathscr{S}_r | v_k \rangle
    &= \sigma_k(f_k^-\tau_k)-(f_k^+\sigma_k)\tau_k \nonumber \\
    &= \nu_k(f_k^- - f_k^+) .
\end{align}
For the normalized eigenvalue $\hat\nu_k=\nu_k/Z$, the spectral beta function
\[
    \beta_k := r\partial_r\log\hat\nu_k
\]
therefore becomes
\begin{equation} \label{eq:spectral_beta_mismatch}
    \boxed{
    \beta_k
    = r\left[
    f_k^- - f_k^+
    -\operatorname{Tr}_{\mathcal T}\left(\frac{P_{\mathcal T}\mathscr{S}_rP_{\mathcal T}}{Z}\right)
    \right] .
    }
\end{equation}
This formula is the main local bridge between microscopic kinematics and the macroscopic spectral RG flow. It shows that the spectral beta function is driven not by the absolute forward--backward mismatch, but by the \emph{centered} mismatch after subtracting the target-active background expansion.

At a fixed point, the condition $\beta_k=0$ imposes
\begin{equation}
    f_k^- - f_k^+
    =
    \operatorname{Tr}_{\mathcal T}\left(\frac{P_{\mathcal T}\mathscr{S}_rP_{\mathcal T}}{Z}\right)
    \qquad \text{for the relevant isolated modes.}
\end{equation}
Equivalently,
\[
(f_k^- - f_k^+) - \langle f^- - f^+\rangle_{\mathcal T}=0 .
\]
The physical meaning is that feature learning stabilizes when the pullback--push-forward entanglement created by gradient descent has no residual shape-changing mismatch in the target-active sector. The incoming and outgoing metric strains need not vanish; their differential, target-visible component must become irrelevant under coarse graining.

\section{Conclusion}
We have argued that the capacity of deep networks to learn hierarchical features can be understood as a geometric phase transition: a target-induced anisotropic instability of a neutral bidirectional learning channel. By exposing the induced Gram metrics and mapping them to a depth-wise RG flow, we obtain an analytical bridge between microscopic parameter updates and the macroscopic, heavy-tailed geometry of deep learning.

The key mechanism is the conversion of asynchronous weight updates into synchronous metric strains. These strains encode a local pullback--push-forward entanglement: the forward leg measures state survival, the backward leg measures gradient visibility, and the target residual couples them. Their target-active symmetric traceless components act as curvature-like sources for the normalized lifted channel. The Callan--Symanzik beta functions measure whether these sources change the shape of the channel or merely rescale it.

The resulting picture is deliberately progressive. We first analyze generic anisotropic generators $\omega_F,\omega_B$ and the RG consequences of their insertions. We then identify their microscopic origin in the synchronous strains produced by gradient descent. In this form, the spectral beta function becomes the centered forward--backward strain mismatch. Heavy-tailed spectra emerge, under the scale-free counting hypothesis, when this centered mismatch becomes irrelevant under depth coarse-graining and the target-active lifted geometry becomes self-similar.

This leaves a clear next step. The present work identifies the local kinematic source of the tensorial
beta functions. A full open-tensor DMFT, which we leave to future work,
should determine self-consistently the statistics of these sources, their
depth correlations, and the emergent pullback--push-forward entanglement
of the target-active lifted spectrum.

\clearpage
\appendix

\section{Tensorial and Spectral Beta Functions}
\label{app:beta_functions}
In this appendix, we explicitly derive the $\beta$-functions for the normalized channel $\hat{\mathcal{C}}(r)$, demonstrating how the isotropic background scale naturally factors out of the geometric evolution, and how the equations for the spectral modes are obtained.

\subsection{Cancellation of Isotropic Terms}
Recall the continuous Callan-Symanzik flow for the unnormalized bidirectional channel $\mathcal{C}(r) = \Sigma(r) \otimes T(r)$:
\begin{align}
    \frac{\partial \mathcal{C}(r)}{\partial r} &= \Sigma \otimes \left( [\gamma I + \omega_F] T \right) - \left( [\tilde{\gamma} I + \omega_B] \Sigma \right) \otimes T \nonumber \\
    &= (\gamma - \tilde{\gamma}) \mathcal{C} + \left( \Sigma \otimes (\omega_F T) - (\omega_B \Sigma) \otimes T \right)
\end{align}
Let $\Gamma(r) = \gamma(r) - \tilde{\gamma}(r)$ represent the total isotropic scalar rate, and let $\mathscr{S}(r) = \Sigma \otimes (\omega_F T) - (\omega_B \Sigma) \otimes T$ denote the anisotropic source operator. The flow simplifies to:
\begin{equation}
    \frac{\partial \mathcal{C}}{\partial r} = \Gamma(r) \mathcal{C}(r) + \mathscr{S}(r)
\end{equation}
We study the normalized tensor $\hat{\mathcal{C}} = \mathcal{C} / Z$, where $Z = \operatorname{Tr}_{\mathcal T}(\mathcal{C})$. The derivative of the partition trace $Z(r)$ is simply the trace of the flow:
\begin{equation}
    \frac{\partial Z}{\partial r} = \operatorname{Tr}\left(\frac{\partial \mathcal{C}}{\partial r}\right) = \Gamma Z + \operatorname{Tr}_{\mathcal T}(\mathscr{S})
\end{equation}
Applying the quotient rule to compute the flow of the normalized tensor $\hat{\mathcal{C}}$:
\begin{align}
    \frac{\partial \hat{\mathcal{C}}}{\partial r} &= \frac{1}{Z} \frac{\partial \mathcal{C}}{\partial r} - \frac{\mathcal{C}}{Z^2} \frac{\partial Z}{\partial r} \nonumber \\
    &= \frac{1}{Z} \big( \Gamma \mathcal{C} + \mathscr{S} \big) - \frac{\mathcal{C}}{Z^2} \big( \Gamma Z + \operatorname{Tr}_{\mathcal T}(\mathscr{S}) \big) \nonumber \\
    &= \Gamma \hat{\mathcal{C}} + \frac{\mathscr{S}}{Z} - \Gamma \hat{\mathcal{C}} - \hat{\mathcal{C}} \frac{\operatorname{Tr}_{\mathcal T}(\mathscr{S})}{Z}
\end{align}
The isotropic expansion/contraction terms $\Gamma \hat{\mathcal{C}}$ cancel out. Multiplying by $r$ to obtain the tensorial $\beta$-function, we get:
\begin{equation}
    \beta(\hat{\mathcal{C}}) = r \frac{\partial \hat{\mathcal{C}}}{\partial r} = r \left( \frac{\mathscr{S}}{Z} - \hat{\mathcal{C}} \operatorname{Tr}\left(\frac{\mathscr{S}}{Z}\right) \right)
\end{equation}
This shows that the geometric evolution of the learning channel depends on the target-aligned anisotropic breaking ($\mathscr{S}$); if $\omega_F = \omega_B = 0$, then $\mathscr{S} = 0$ and $\beta(\hat{\mathcal{C}}) = 0$.

\subsection{The Spectral Beta Function}
To analyze the critical fixed point, we project the tensorial $\beta$-function onto the target-active eigenmodes. Let $\hat{\nu}_k$ be an eigenvalue of $\hat{\mathcal{C}}$ with corresponding normalized eigenvector $v_k$. As noted in Section VI.A, we focus on isolated target-active modes whose eigengaps are sufficiently large to neglect second-order eigenvector rotations. By standard first-order eigenvalue perturbation (or the Hellmann-Feynman theorem), the derivative of the eigenvalue is the expectation value of the derivative operator:
\begin{align}
    \frac{\partial \hat{\nu}_k}{\partial r} &= v_k^T \left( \frac{\partial \hat{\mathcal{C}}}{\partial r} \right) v_k \nonumber \\
    &= v_k^T \left( \frac{\mathscr{S}}{Z} - \hat{\mathcal{C}} \operatorname{Tr}\left(\frac{\mathscr{S}}{Z}\right) \right) v_k \nonumber \\
    &= \left\langle \frac{\mathscr{S}}{Z} \right\rangle_k - \hat{\nu}_k \operatorname{Tr}\left(\frac{\mathscr{S}}{Z}\right)
\end{align}
where $\langle \cdot \rangle_k$ denotes the projection along the $k$-th mode. The spectral $\beta$-function is defined as the logarithmic derivative of the mode:
\begin{equation}
    \beta_k = r \frac{\partial \log \hat{\nu}_k}{\partial r} = \frac{r}{\hat{\nu}_k} \frac{\partial \hat{\nu}_k}{\partial r}
\end{equation}
Substituting the previous result, we finally obtain:
\begin{equation}
    \beta_k = r \left( \frac{1}{\hat{\nu}_k} \left\langle \frac{\mathscr{S}}{Z} \right\rangle_k - \operatorname{Tr}\left(\frac{\mathscr{S}}{Z}\right) \right)
\end{equation}
At the non-trivial RG fixed point driven by the gradient's dynamic feedback, the network adjusts the anisotropic operator $\mathscr{S}$ such that $\beta_k \approx 0$ for all modes aligned with the target.

\subsection{Kinematic Accumulation of Symmetry-Breaking Operators}
To understand why the continuous symmetry-breaking operators $\omega_F$ and $\omega_B$ asymptote to non-zero target-aligned states, we must trace them back to the exact kinematics of gradient descent. At training step $t$, the bare gradient update for the weight matrix at layer $r$ is:
\begin{equation}
    \Delta W^{(r)}(t) = - \frac{2\eta}{M} \sum_{\mu=1}^M r^\mu(t) c_r^\mu (u_{r-1}^\mu)^T
\end{equation}
where $c_r^\mu = A_r^\mu b_r^\mu$ is the masked conjugate field. Notice a crucial structural distinction in the indices. The weight update $\Delta W^{(r)}$ is driven by an \textit{asynchronous} outer product $c_r^\mu (u_{r-1}^\mu)^T$, bridging the input space and the output space of the layer. However, the resulting geometric metric deformations resolve into \textit{synchronous} outer products.

To see this, we analyze the evolution of the forward metric $(J_r)^T J_r = (W^{(r)})^T A_r W^{(r)}$, which acts entirely within the input space, inside a fixed activation chamber. At first order, its variation is:
\begin{equation}
    \Delta \left[ (J_r)^T J_r \right] \approx (\Delta W^{(r)})^T A_r W^{(r)} + (W^{(r)})^T A_r \Delta W^{(r)}
\end{equation}
By substituting the gradient and utilizing the backward transport identity $(b_{r-1}^\mu)^T = (c_r^\mu)^T W^{(r)}$, the chain rule naturally pulls the backward field down to layer $r-1$, aligning the indices synchronously. The exact variation becomes:
\begin{equation}
    \Delta \left[ (J_r)^T J_r \right] = - \frac{2\eta}{M} \sum_{\mu=1}^M r^\mu(t) \left[ u_{r-1}^\mu (b_{r-1}^\mu)^T + b_{r-1}^\mu (u_{r-1}^\mu)^T \right]
\end{equation}
Dually, the backward visibility metric $J_r (J_r)^T = A_r W^{(r)} (W^{(r)})^T A_r$ acts on the output space. Its variation pushes the forward field up to layer $r$ using $u_r^\mu = A_r W^{(r)} u_{r-1}^\mu$, yielding synchronous output-space terms:
\begin{equation}
    \Delta \left[ J_r (J_r)^T \right] = - \frac{2\eta}{M} \sum_{\mu=1}^M r^\mu(t) \left[ c_r^\mu (u_r^\mu)^T + u_r^\mu (c_r^\mu)^T \right]
\end{equation}

The total metric at an asymptotic time $T$ is the temporal integral of these exact deformations, added to the random isotropic initialization $\tau_r I$ and $\tau_r A_r$. Thus, the symmetry-breaking operators $\Omega_r^F$ and $\Omega_r^B$ (which map to $\omega_F(r)dr$ and $\omega_B(r)dr$ in the continuous limit) are formally defined as the integrated accumulation of these kinematic synchronous updates:
\begin{align}
    \Omega_r^F &\equiv - \frac{2\eta}{\epsilon M} \sum_{t=0}^T \sum_{\mu=1}^M r^\mu(t) \left[ u_{r-1}^\mu(t) (b_{r-1}^\mu(t))^T + \text{h.c.} \right] \\
    \Omega_r^B &\equiv - \frac{2\eta}{\epsilon M} \sum_{t=0}^T \sum_{\mu=1}^M r^\mu(t) \left[ c_r^\mu(t) (u_r^\mu(t))^T + \text{h.c.} \right]
\end{align}
where $\text{h.c.}$ denotes the Hermitian conjugate. Because gradient descent consistently aligns the forward and backward fields with the target $Y$ over time, the residual $r^\mu(t)$ does not oscillate randomly along the active task directions. Consequently, these integrals need not vanish; under coherent target alignment they can accumulate a permanent geometric deformation.

\subsection{A Consistent Non-Trivial Fixed-Point Construction}
To demonstrate that $\beta_k = 0$ admits physically meaningful solutions where the symmetry breaking is non-zero ($\omega_F, \omega_B \neq 0$), we construct an explicit asymptotic solution based on the kinematic accumulations derived above. Recall the spectral fixed-point condition for a mode $v_k = \psi_B \otimes \psi_F$:
\begin{equation} \label{eq:beta_zero}
    \frac{\langle v_k | \mathscr{S} | v_k \rangle}{\nu_k} = \frac{\operatorname{Tr}_{\mathcal T}(\mathscr{S})}{Z}
\end{equation}
This equation dictates that at the critical point, the local relative anisotropic gain of the target mode must match the global average anisotropic gain of the entire tensor trace.

When the network reaches its asymptotic late-training basin, the error residual $r(t)$ decays to zero, and the instantaneous gradient updates vanish ($\Delta W \to 0$). However, as shown in Appendix A.3, the symmetry-breaking operators $\omega_F$ and $\omega_B$ represent the \textit{integrated accumulation} of these updates over the entire training trajectory. Because gradient descent consistently pushes the weights in the direction of the target, the integral of this target-directed pulse asymptotes to a non-zero, rank-one perturbation aligned with the target state:
\begin{align}
    \omega_F &\to \omega_* |\psi_F\rangle \langle \psi_F| \\
    \omega_B &\to \tilde{\omega}_* |\psi_B\rangle \langle \psi_B|
\end{align}
where $\omega_*, \tilde{\omega}_* > 0$ are the frozen asymptotic magnitudes of the accumulated anisotropies.

We evaluate the local gain on this target-aligned mode $v_k$. By the definition of $\mathscr{S} = \Sigma \otimes (\omega_F T) - (\omega_B \Sigma) \otimes T$, the projection is:
\begin{align}
    \langle v_k | \mathscr{S} | v_k \rangle &= \langle \psi_B | \Sigma | \psi_B \rangle \langle \psi_F | \omega_F T | \psi_F \rangle \nonumber \\
    &- \langle \psi_B | \omega_B \Sigma | \psi_B \rangle \langle \psi_F | T | \psi_F \rangle
\end{align}
Substituting the rank-one asymptotic ansatz, we have $\langle \psi_F | \omega_F T | \psi_F \rangle = \omega_* \langle \psi_F | T | \psi_F \rangle$ and $\langle \psi_B | \omega_B \Sigma | \psi_B \rangle = \tilde{\omega}_* \langle \psi_B | \Sigma | \psi_B \rangle$. Thus:
\begin{equation}
    \langle v_k | \mathscr{S} | v_k \rangle = \nu_k (\omega_* - \tilde{\omega}_*)
\end{equation}
Substituting this back into Eq. \ref{eq:beta_zero}, the $\nu_k$ cancels out, and the fixed-point condition reduces to a simple scalar algebraic equation:
\begin{equation} \label{eq:asymptotic_balance}
    \omega_* - \tilde{\omega}_* = \frac{\operatorname{Tr}_{\mathcal T}(\mathscr{S})}{Z}
\end{equation}

This provides an explicit mean-field construction of a non-trivial fixed point. The equation expresses a dynamic geometric balance: the network halts its geometric evolution when the net difference between the accumulated forward pull ($\omega_*$) and the backward push ($\tilde{\omega}_*$) equals the trace-diluted average of the entire anisotropic operator. Because $\omega_*$ and $\tilde{\omega}_*$ are generated dynamically by independent forward and backward signal propagation, the system can possess the degrees of freedom required to satisfy this algebraic constraint with non-zero asymptotic couplings.

\end{document}